\documentclass[aps, %
 reprint,
 amsmath,amssymb,
 prl,
]{revtex4-2}
\usepackage{graphicx}
\usepackage{dcolumn}
\usepackage{color}
\usepackage{bm}
\usepackage{siunitx}
\newcommand{\ket}[1]{\left\vert #1\right\rangle}

\newcommand{\brakets}[3]{\left\langle #1\left\vert#2\right\vert#3\right\rangle}
\newcommand{\ii}{\mathrm{i}}
\newcommand{\ee}{\mathrm{e}}
\newcommand{\dd}{\mathrm{d}}
\DeclareUnicodeCharacter{2061}{}
\begin{document}
\preprint{APS/123-QED}
\title{Nanoscale Femtosecond Coherent Radiation and Spatiotemporally Shaped free electron Wavefunction}
\author{Wu Wen$^{1,2}$}
\author{Jing Li$^{1,2}$}
\author{Yunquan Liu$^{1,2,3,4}$}
\email{Yunquan.liu@pku.edu.cn}

\affiliation{%
$^1$State Key Laboratory for Mesoscopic Physics and Collaborative Innovation Center of Quantum Matter,\\
School of Physics, Peking University, Beijing 100871, China\\
$^2$National Biomedical Imaging Center, Peking University, Beijing 100871, China\\
$^3$Collaborative Innovation Center of Extreme Optics, Shanxi University, Taiyuan, Shanxi 030006, China\\
$^4$Peking University Yangtze Delta Institute of Optoelectronics, Nantong, Jiangsu 226010, China}

\date{\today}

\begin{abstract}
We study tunable nanoscale femtosecond coherent radiation based on a coupled nanowire pair (CNP) structure that is excited by a strong laser. The structure functions as a nanoscale undulator (NU): the electrons moving through the nanogap are driven by a spatially periodic, transverse optical near-field. We show that the transverse near-field can {actively} shape the electron wavefunction by inducing both a periodic oscillation and a quantum squeezing of its width. We then validate this theoretical framework {by numerically solving the relativistically corrected time-dependent Schrödinger equation (RC-TDSE)}. The generated femtosecond pulse trains can be spectrally, temporally, and spatially controlled. 
This framework establishes the transverse optical near-field interaction as a novel mechanism to spatiotemporally shape electron wavefunctions, which illuminates a path to versatile platform for on-chip femtosecond coherent light source and the application in free-electron quantum optics.
\end{abstract}

\maketitle

Interactions between free electrons and tailored optical fields in nanophotonic structures offer versatile platforms \cite{OpticalExcitationsElectron2021,garciadeabajoOpticalExcitationsElectron2010}. Recent significant progress includes the development of dielectric laser accelerators and a variety of free electron-driven light sources \cite{Talebi2019NatCommun,chloubaCoherentNanophotonicElectron2023a,wongGraphenePlasmonbasedFreeelectron2016a,zhangCoherentSurfacePlasmon2022a,Taleb2023Nat.Phys.,Taleb2025NatCommun}. Many of free electron-driven light sources rely on spontaneous emission processes, such as Smith-Purcell radiation \cite{yangPhotonicFlatbandResonances2023}, Cherenkov radiation \cite{linBrewsterRouteCherenkov2021}, and transition radiation \cite{chenLowVelocityFavoredTransitionRadiation2023}.
{Stimulated interactions have garnered significant interest since the advent of ultrafast electron microscopy (UEM)} \cite{gahlmannUltrashortElectronPulses2008,yangScanningUltrafastElectron2010a,zewailFourDimensionalElectronMicroscopy2010,OpticalExcitationsElectron2021}. {A} prominent application is photon-induced near-field electron microscopy (PINEM)  \cite{barwickPhotoninducedNearfieldElectron2009,parkPhotoninducedNearfieldElectron2010}, which {exploits} the interaction of free electrons with the longitudinal component of the optical near-field. This leads to quantum-coherent modulation of the wavefunction \cite{feistQuantumCoherentOptical2015}.
PINEM has inspired wide applications in shaping electron wavefunctions \cite{feistQuantumCoherentOptical2015,Talebi2016NewJ.Phys., ebelInelasticElectronScattering2023,LiPhysRevLett2024}, ultrafast dynamical imaging \cite{kurmanSpatiotemporalImaging2D2021}, and quantum optics \cite{khalafComptonScatteringDriven2023}. However, the efficient generation of tunable, femtosecond coherent radiation \cite{dangDeepsubcycleUltrafastOptical2024}, particularly through active spatiotemporal shaping electron wavefunction by transverse {optical} near-fields, remains a significant open challenge.

{Undulators, which induce periodic oscillations in relativistic charged particles, are the foundation of free electron light sources such as synchrotrons and free electron lasers (FELs) \cite{Elder1947Phys.Rev.,Elias1976Phys.Rev.Lett.,Pellegrini2016Rev.Mod.Phys.}. Conventional designs, based on centimeter-scale periodic magnetic fields from permanent-magnet or superconducting arrays, require large facilities and high-energy electron beams \cite{Kumakhov1976PhysicsLettersA,Halbach1980NuclearInstrumentsandMethods}. This has motivated intensive research into compact alternatives. Crystalline undulators, for instance, use the atomic-lattice potential to generate MeV-scale photons from channeled particles \cite{Wistisen2014Phys.Rev.Lett.,Bellucci2004Phys.Rev.STAccel.Beams}. A particularly promising approach is the dielectric-laser undulator, in which the optical-period laser field yields high-energy photons from significantly lower-energy electrons \cite{Milburn1963Phys.Rev.Lett.,Plettner2008Phys.Rev.STAccel.Beams,Schmid2022Phys.Rev.Accel.Beams}. This process is understood as inverse Compton scattering (ICS). The advent of nanophotonics further enables structured optical near fields—photonic quasiparticles \cite{riveraLightMatterInteractions2020b}—to act as an undulator, offering extreme miniaturization and unprecedented quantum control of the emitted radiation.}


In this Letter, we theoretically propose and numerically demonstrate a novel scheme to address this challenge by introducing a coupled nanowire pair (CNP) structure \cite{akimovGenerationSingleOptical2007,fangPlasmonicCouplingBow2011,xuSinglemodeLasingGaN2012,huangModelingEvanescentCoupling2007,yanDirectPhotonicPlasmonic2009} as a “nanoscale undulator (NU)". 
NU is realized with the localization-enhanced, spatially periodic transverse electric near-fields within the CNP's nanogap excited by an external pump laser. {In the context of ICS in a NU, we treat the scattering events within a quantum electrodynamical (QED) framework. }
By numerically solving the relativistically corrected time-dependent Schrödinger equation (RC-TDSE) \cite{Talebi2016NewJ.Phys., ebelInelasticElectronScattering2023,talebiStrongInteractionSlow2020a}, we unveil the dynamics of how the electron wavefunction is spatiotemporally shaped within this NU. We show that the shaped electron simultaneously emits tunable, nanoscale, and femtosecond coherent radiation. The incident electron kinetic energy can be used to control the wavelength and pulse length of the radiation. This work elucidates the quantum dynamics of free electrons in strong, confined transverse optical fields \cite{garciadeabajoOpticalModulationElectron2021,kozakElectronVortexBeam2021,tsessesTunablePhotoninducedSpatial2023,kozakInelasticPonderomotiveScattering2018,madanUltrafastTransverseModulation2022} and leverages this dynamics to realize a versatile platform for a new class of on-chip coherent light sources \cite{adamoLightWellTunable2009,wangNovelLightSource2017}.

The proposed NU is schematically illustrated in Fig. 1(a). It consists of a CNP formed by two nanowires \cite{wuPhotonicNanolaserExtreme2022}. Each nanowire has a cross section with a side length $R= \qty{138} {nm}$ and a nanowire length $L=\qty{5}{\mu m}$. The two nanowires are separated by a 10 nm gap. In the model, the nanowire material is chosen to be lossless with a refractive index of 2. An incident free electron travels along the $z$-axis in the $x-z$ plane $(y=0)$.

The wavevector of the $\qty{800}{nm}$ $x-$linearly polarized pump laser with $1.5\times10^9$ V/m amplitude is perpendicular to the $x-z$ plane. A few low-order TE-like gap modes within the CNP could be excited by the pump laser. The finite-difference time-domain (FDTD) simulation (Fig. 1) {reveals} that the excitation primarily produces a strong transverse electric-field component, $E_x$, within the gap. The simulation further shows that the $E_x$ {varies periodically along} $z$ with propagation constant $k_p=1.3\,k_0$ (where $k_0$ is the {pump's} free-space wavevector), realizing the NU’s periodic deflecting field. As shown in Fig. 1(b), the simulated peak-to-background field intensity ratio (gap mode $vs.$ field inside nanowires) is $\sim 10$. {By symmetry}, the $E_y$ and $E_z$ on the electron's trajectory plane ($y=0$) are negligible except {near} the two end facets, confining electron's motion to the $x-z$ plane.

\begin{figure}
\includegraphics[width=1\columnwidth]{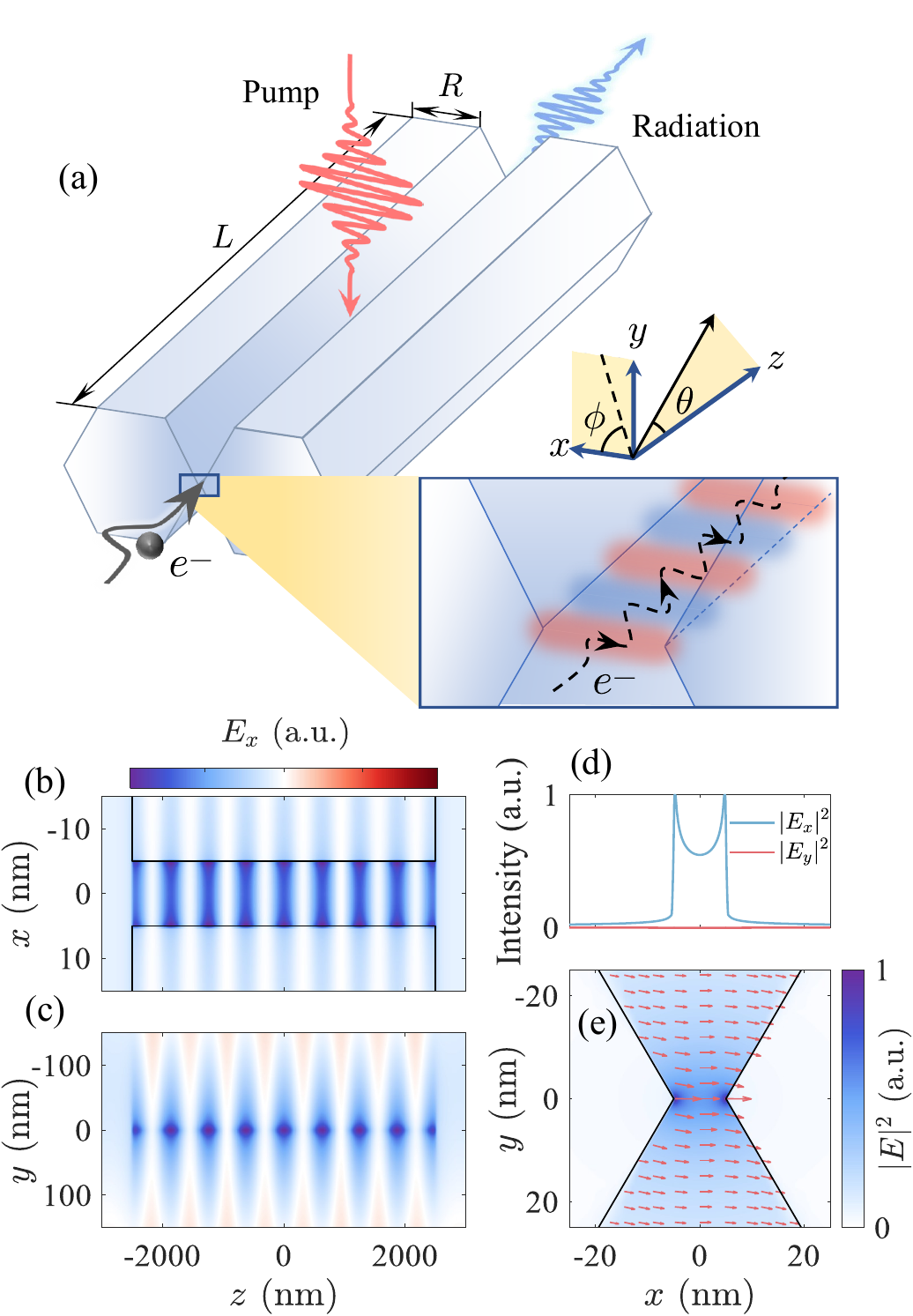}
\caption{\label{fig:1} NU based on a CNP for coherent radiation generation. (a) Schematic of the proposed set-up. An electron propagates through the nanogap (10 nm width) between two parallel hexagonal nanowires ($L=\SI{5}{\mu m}$). The inset illustrates the electron's transverse oscillation trajectory within the gap, driven by the gap mode. {(b) Simulated $E_x$ field distribution in the $x-z$ cross-section of the CNP gap with $y=0$, excited by an 800 nm linearly polarized laser with the amplitude of $1.5\times10^9$ V/m along $x$. The black solid lines indicate the nanowire boundaries. (c) the $E_x$ field distribution in the $y-z$ cross-section with $x=0$. (d) The field intensity within CNP gap. The sampled line is $y=0$ in the cross-section shown in (e). (e) The electric field distribution in the $x-y$ cross-section with $z=0$. The arrows represent the directions and amplitudes of the electric field.}}
\end{figure}

We develop an analytical model for the interaction between a free electron and NU modes \cite{SM}. {Including relativistic corrections, the minimal-coupling Hamiltonian in the Coulomb gauge for a single electron with initial energy} $E_e$, charge $e$, and rest mass $m_e$ {reads} \cite{eldarSelfTrappingSlowElectrons2024}:
\begin{equation}
\begin{aligned}
    \hat{H}&= E_e + \bm{v}_e \cdot (\hat{\bm{p}} - \bm{p}_0) + \frac{(\hat{\bm{p}}_\perp - e\bm{A}_\perp)^2}{2\gamma m_e}\\&+ \frac{(\hat{p}_z - p_0 - eA_z)^2}{2\gamma^3 m_e} + e\phi_e,
\end{aligned}
\end{equation}
where $v_e$, $\gamma$, $\bm{p}_0$, $\hat{p}_z$, and $\hat{\bm{p}}_\perp$ are the initial velocity, Lorentz factor, initial momentum, longitudinal and transverse momenta of the electron, respectively. {We denote the NU-mode scalar potential by} $\phi_e$, and {decompose the vector potential $\bm{A}$ into longitudinal and transverse components, $A_z$ and $\mathbf{A}_{\perp}$, respectively}. {For analytic tractability, we model one of the NU near field modes as} $E_x=E_0\cos(k_pz-\omega_pt)\cosh(\beta x)$ in the subsequent calculation, where $E_0$ is the field amplitude, $\omega_p$ and $k_p$ are the {mode} frequency and longitudinal wavevector, {and} $\beta$ {sets} the {evanescent decay along} $x$ (decay length $\beta^{-1}$).

The initial electron wavefunction is taken as a Gaussian packet with standard deviations $\sigma_{z0}$ and $\sigma_{x0}$ {along} $z$ and $x$, centered at $x=0$. The interaction induces transverse “wiggling", {the} operating principle of {the} NU. The wave packet center {then} follows the closed-form (see the supplemental materials \cite{SM}):
\begin{equation}
    x_c(z,t) = \frac{eE_0}{\gamma m_e \omega_p \Omega} \left. \cos[k_p(z-v_et) - \Omega\tau] \right|_{\tau=\tau_1}^{\tau=\tau_2}.
\end{equation}
Here, $\Omega=v_ek_p-\omega_p$ is the phase-mismatch frequency, which also {sets} the electron's transverse oscillation frequency. The terms $\tau_1=\max\{0,t-(z+L/2)/v_e\}$ and $\tau_2=\min\{t,t-(z-L/2)/v_e\}$ {bound the interaction window for an electron at position $z$ and time $t$.} After interacting with the NU, the electron's wavefunction acquires a periodic transverse modulation with {spatial period} $\lambda_e=2\pi v_e/\omega_p$. {This oscillation establishes the NU’s fundamental feasibility.}

A long interaction length is essential for high radiation efficiency but is fundamentally limited by lateral dispersive broadening of the electron wave packet. {Because of the uncertainty principle, the tight nanoscale confinement induces large momentum uncertainty and rapid dispersion} —a ubiquitous challenge in free-electron control. The proposed NU {intrinsically mitigates this}: the evanescent gap mode {supplies} a strong, {near}-harmonic transverse potential that squeezes the wave packet. For an initially unchirped Gaussian packet, this squeezing preserves {its} profile. The transverse width $\sigma_x\left(z,t\right)$ {evolves as} \cite{SM}:
\begin{equation}
    \begin{aligned}
        \sigma_x(z,t) &= \sigma_{x0} \sqrt{\cos^2(\omega t') + M \sin^2(\omega t')} \;,\\
        M &= \frac{\hbar^2}{4\gamma^2m_e^2\omega^2\sigma_{x0}^4}.
    \end{aligned}
\end{equation}
Here, $t'$ is the interaction duration, and $M$ is a dimensionless squeezing parameter. The {width oscillation frequency}, $\omega$ is set by the strength of the {effective harmonic confinement}, which {scales with} $E_0$ and $\beta$. Squeezing occurs {for} $M<1$ {, achieved when the field amplitude exceeds} $E_\mathrm{th}=\hbar\omega_p/(2\sqrt{2\gamma}e\beta\sigma_{x0}^2)$. In this regime, the wave packet is compressed $\left(\sigma_x\left(t\right)<\sigma_{x0}\right)$, overcoming its natural dispersion. The coherent squeezing {enables} long-range energy transfer and radiation in the NU.

The net energy {exchange} between the electron and the near field {is} characterized by the electron energy loss spectrum (EELS). EELS is governed by the phase imprinted on the electron wavefunction, and arises from two dominant contributions. {First}, a ponderomotive phase modulation induced by the transverse near field, with strength quantified by the coupling parameter $g\propto e^2E_0^2 /(8\gamma m_e\hbar\omega_p^2\Omega)$. 
Here, $g$ {increase} as the electron velocity $v_e$ approaches the phase-matching condition ($\Omega\to0$), which produces more prominent sidebands in EELS that reflect multiphoton net exchange. Second, a dynamical phase generated by the wave packet’s transverse squeezing motion. This two-part structure is unique to transverse-field interactions, which distinguishes our mechanism from conventional longitudinal-field PINEM. Consequently, EELS is a ponderomotive modulation “dressed” by an additional, squeezing-induced dynamical phase.

\begin{figure*}
\includegraphics[width=2\columnwidth]{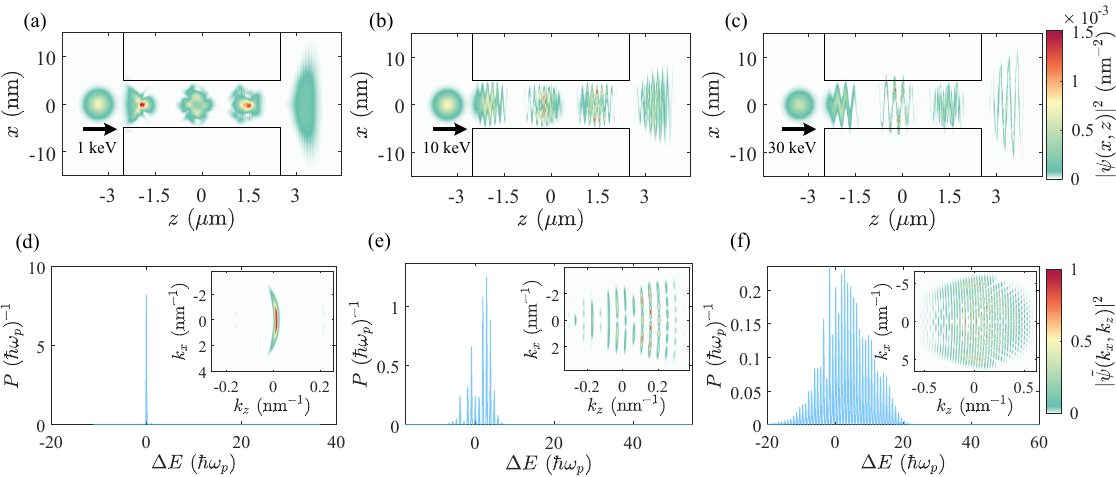}
\caption{\label{fig:2} Electron wavefunction evolution and EELS for different incident kinetic energies (1 keV, 10 keV, and 30 keV) in the NU. (a)-(c) Snapshots of the real-space electron probability density $|\psi (x,z,t)|^2$ during interaction (the time evolutions of the wave packets are provided in Supplemental Material movie). (d)-(f) Corresponding EELS spectra $P(\Delta E)$ after interaction. Discrete peaks separated by a single photon energy are observed. Insets: Final probability density in the momentum-space $|\tilde{\psi}(k_x, k_z)|^2$.}
\end{figure*}

To validate the analytical model, we numerically solved RC-TDSE using the near-fields generated from our FDTD simulations \cite{SM}. We first consider a $\SI{1}{keV}$ incident electron. Figure 2(a) shows snapshots of the electron probability density $|\psi(x, z, t)|^2$ {that highlight} the two key predicted effects. First, the transverse near field drives a periodic oscillation of the wave packet centroid with spatial period of $\sim$49 nm, in excellent agreement with the theoretical $\lambda_e$. Second, the simulation confirms coherent squeezing: the field amplitude in our structure ($E_0\approx\SI{1e10}{V/m}$) significantly exceeds the theoretical threshold $E_\mathrm{th}$, providing strong confinement that overcomes natural dispersion, {and keeps} the wave packet remains localized within the gap throughout the interaction. These results {show} that the CNP structure {supports} a stable, long-range interaction.

We next examine the tunability {at} higher incident kinetic energies of $10$ keV and $30$ keV. The real-space dynamics in Figs. 2(a-c), reveal markedly stronger transverse oscillation for higher-energy electrons. This trend agrees with the analytical model [Eq. (2)]{, e.g., the oscillation amplitude is inversely proportional to the phase-mismatch frequency} $\Omega$. Specifically, as the kinetic energy increases from 1 keV to 30 keV, $\Omega$ decreases from {\SI{345}{THz} to $\SI{218}{THz}$, directly enhancing transverse modulation. The EELS is calculated as \cite{talebiStrongInteractionSlow2020a}:
\begin{equation}
    P(E)=\frac{\mathrm{d}}{\mathrm{d}E}\langle\psi(x,z,t_f)|\hat{T}|\psi(x,z,t_f)\rangle,
\end{equation}
where $\hat{T}$ is the kinetic energy operator and $t_f$ is the time after the interaction. The calculated EELS and the momentum-space probability densities{,} $(|\tilde{\psi}\left(k_x, k_z\right)|^2)$, for the {\SI{1}{keV}}, {\SI{10}{keV}}, and {\SI{30}{keV}} appear in Figs. 2(d-f). The results of the calculated EELS match the analytical expectations. Comparison across energies confirms that the interaction strength increases with the electron energy, as evidenced by (i) broader EELS sidebands, indicating larger net energy exchange, and (ii) more complex momentum-space structure with larger transverse spread, signifying stronger modulation. Furthermore, the EELS is notably asymmetric{—a signature of nonadiabatic longitudinal fields localized at the NU’s entrance and exit facets.} These fields impart sharp momentum kicks to the electron, and the interference between these boundary effects and the continuous {gap} interaction {sets} the final asymmetric profile \cite{SM}.

The transversely oscillating, quantum-squeezed electron wave packet acts as a nanoscale coherent light source. Fundamentally, the emission can be treated as a second-order scattering process between the electrons and the NU near-field photons (or quasi-particles) \cite{riveraLightMatterInteractions2020b,khalafComptonScatteringDriven2023}. {We can describe the radiation in a QED framework. The radiation field is treated as a quantized field with the form $\hat{\bm{E}_e}(r)=\sum_{\lambda}(\bm{E}_\lambda\hat{a}_\lambda+\bm{E}^*_\lambda\hat{a}^\dagger_\lambda)$ while the driven laser is retained as classical field. The Hamiltonian of the interaction of the electron and quantized field is $\hat{H}_{\mathrm{I}}=-\hat{\bm{A}}_e\cdot\hat{\bm{j}}_t$.
Here $\hat{\bm{A}}_e$ is the vector potential of the quantized radiation field. The current density operator $\hat{\bm{j}}_t$ is defined as:
\begin{equation}
    \hat{\bm{j}}_t=\frac{e}{\gamma m_e}\bm{\Gamma}\cdot(\hat{\bm{p}}-e\bm{A})+e\bm{v}_e\frac{v_e^2}{c^2},
\end{equation}
where $\bm{\Gamma}=\mathrm{diag}(1,1,\gamma^{-1})$.
Since the emitted photon energy is on the order of eV, the quantum recoil from the emitted photons is neglected. In this limit, the electron current drives each radiation mode $\lambda$ into a coherent state, with a
complex amplitude:
\begin{equation}
    \alpha_\lambda(t)=\frac{1}{\hbar\omega_\lambda}\int_{t_0}^t\dd\tau~\ee^{\ii\omega_\lambda(\tau-t_0)}\brakets{\psi_e(\tau)}{\bm{E}_\lambda^*\cdot\hat{\bm{j}}_t}{\psi_e(\tau)},
\end{equation}
where $\ket{\psi_e(\tau)}$ is the electron state and $\omega_\lambda$ is the eigen frequency of mode $\lambda$. 
Crucially, the expectation value of the total quantized field evolves identically to a classical field. 
Therefore, we calculate the radiation intensity by numerically solving Maxwell's equations with the quantum current density expectation value, $\brakets{\psi_e(\tau)}{\hat{\bm{j}}_t}{\psi_e(\tau)}$.}



\begin{figure}
\includegraphics[width=1\columnwidth]{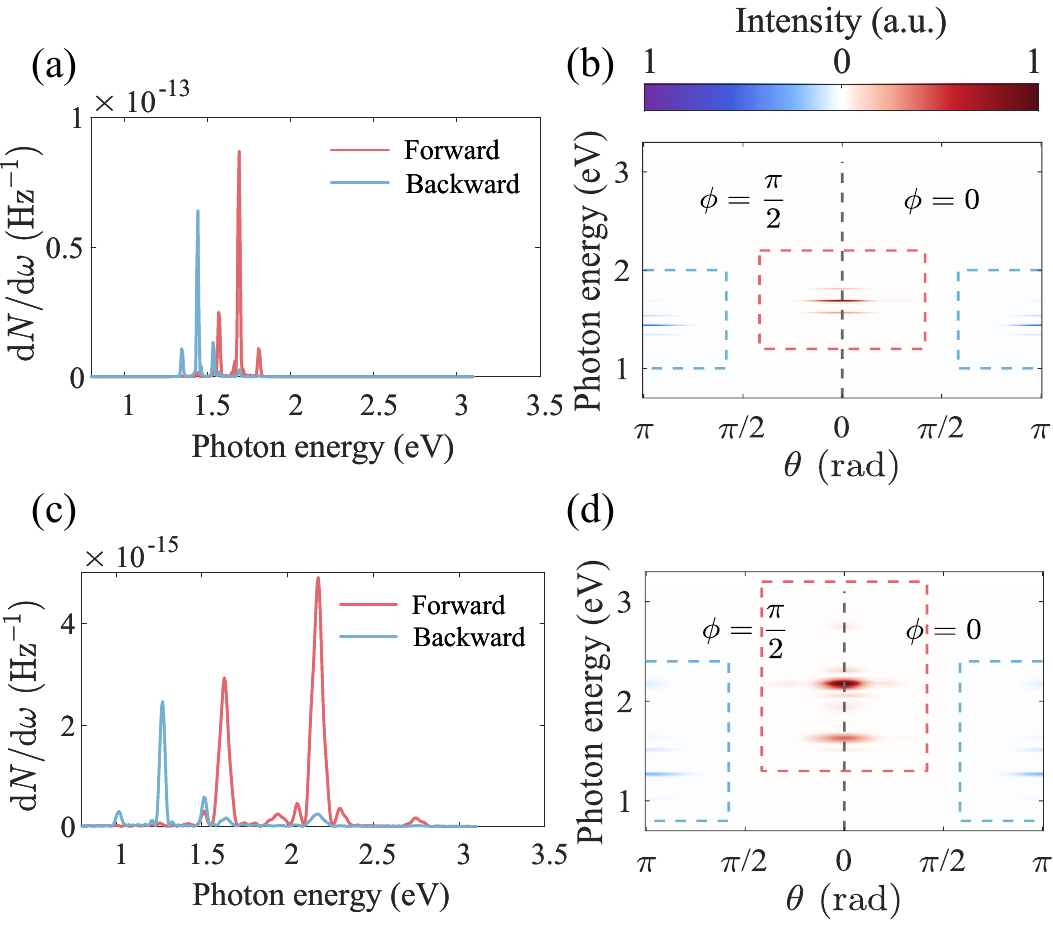}
\caption{\label{fig:3} Spectrum and spatial distribution of the radiation from actively shaped electrons. (a), (c) Radiation spectra for electrons with energies of $\SI{1}{keV}$ and $\SI{10}{keV}$, respectively. (b), (d) Corresponding normalized spatial distributions of the radiation intensity. The polar ($\theta$) and azimuthal ($\phi$) angles are defined in Fig. 1. Forward and backward radiation are distinguished by color and dashed line box.}
\end{figure}

The photon energy is determined by the electron’s oscillation frequency $\Omega$, and the relativistic Doppler effect. For an emission angle $\theta$, the fundamental harmonic appears at $\omega_r(\theta)= \Omega/(1 - v_e\cos\theta/c)$ \cite{jackson_classical_1999}. {The NU near field can be decomposed into three dominant spatial-frequency components}: forward-propagating ($k_p$), the backward-propagating ($-k_p$), and the non-propagating (NP, $k_z \approx 0$). The electron interacts with {all three} simultaneously, {yielding} effective oscillation frequencies ($\Omega_{\mathrm{ICS}},\,\Omega_{\mathrm{CS}},\,\Omega_{\mathrm{NP}}$). Consequently, three prominent peaks {would} appear in the photon energy spectrum, corresponding to three underlying processes \cite{khalafComptonScatteringDriven2023}: (i) Inverse Compton Scattering (ICS) from the co-propagating field component \cite{dangDeepsubcycleUltrafastOptical2024}; (ii) Compton Scattering (CS) from the counter-propagating component; and (iii)  emission from the NP evanescent component.

Considering the QED effect, we show the calculated radiation spectrum from a 1 keV electron with $\sigma_{z0}=35$ nm, integrated over the forward ($\theta<\pi/2$) and backward ($\theta>\pi/2$) hemispheres in Fig. 3(a). As predicted, each direction exhibits three distinct peaks in the forward direction at 1.54, 1.69, and 1.81 eV and in the backward direction at 1.34, 1.44, and 1.53 eV.} These radiation corresponds to the  effective oscillation frequencies of ($\Omega_{\mathrm{ICS}},\,\Omega_{\mathrm{CS}},\,\Omega_{\mathrm{NP}}$), confirming that the radiation originates from the electron's simultaneous interaction with the forward-propagating, backward-propagating, and NP near-field components. The spatial distribution further characterizes the emission profile,  as seen in Fig. 3(b).

One of key advantages of the NU is its frequency tunability via the incident electron's kinetic energy, as seen by comparing the 1-keV [Fig. 3(a)] and 10-keV [Fig. 3(c)] spectra. Increasing the energy significantly blueshifts the forward radiation peaks while redshifting the backward radiation. This tuning is direct consequence of the energy-dependence on both $\Omega$ and the relativistic Doppler factor. The 10 keV spectra also exhibit more pronounced broadening. Additionally, the NU affords excellent control over the spatial and angular properties of the radiation: the angular distributions [Figs. 3(b) and 3(d)] show extremely low spatial dispersion for each spectral peak. The emission shows negligible angular dispersion. This inherent modal selectivity and a stable, high-quality output beam suited for integration and practical applications.

\begin{figure}
\includegraphics[width=1\columnwidth]{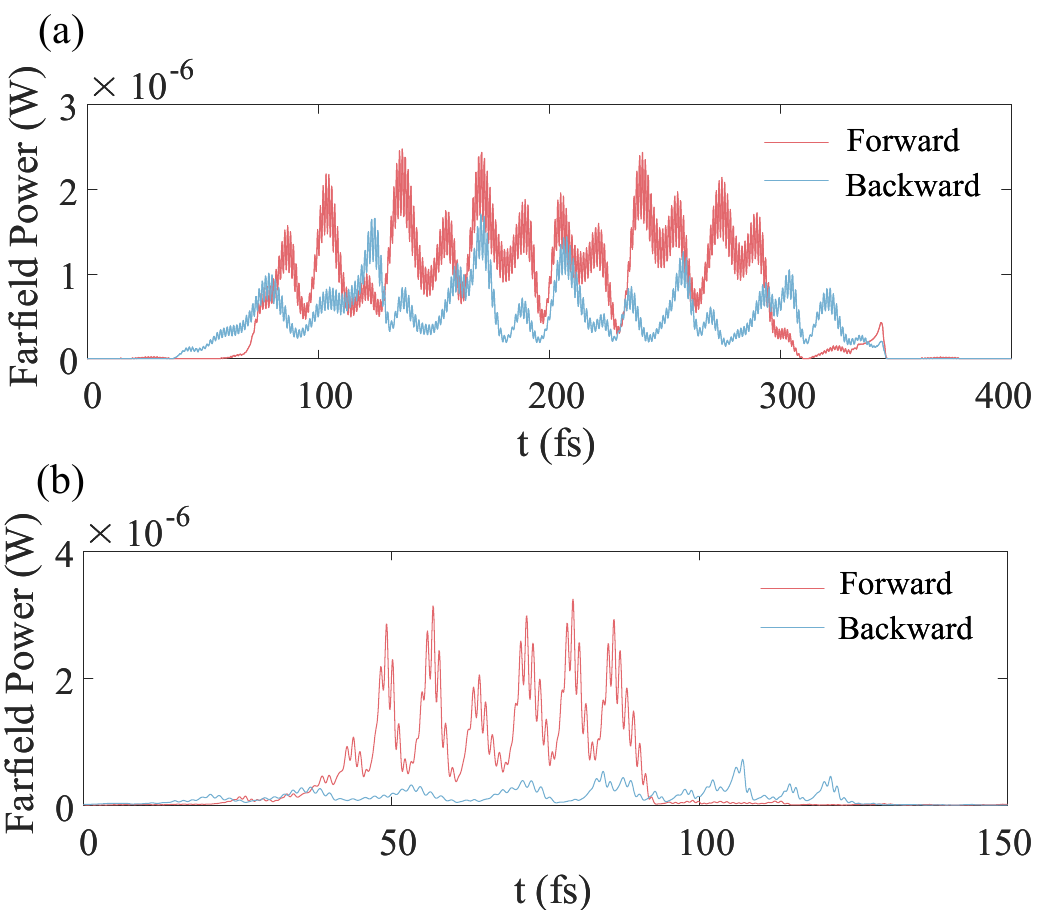}
\caption{\label{fig:4} Farfield power of forward and backward radiation for a free electron with an energy of (a) $\SI{1}{keV}$ and (b) $\SI{10}{keV}$.}
\end{figure}

The spatially {periodicity} of the NU near field directly imprints temporal structure on the radiation. As the electron transverses the structure, it experiences a near field that varies with a period $\lambda_p=2\pi/k_p$. Emission is strongest at the field antinodes, producing a sequence of bursts that naturally forms an equally spaced train of femtosecond pulses. For forward radiation, the pulse period $T_f$ is set by the different time interval to pass one NU period between the electron and the radiation. Estimating the travel time over one field period $\lambda_p$ yields:
\begin{equation}
    T_f \approx \frac{\lambda_p}{v_e}-\frac{\lambda_p}{c_g}=\frac{2 \pi}{k_p}\left(\frac{1}{v_e}-\frac{1}{c_g}\right),
\end{equation} where $c_g$ is the group velocity of the radiated pulse within the CNP.

Our simulations confirm this temporal imprinting. As shown in Fig. 4(a), the forward emission for 1 keV electron forms a femtosecond pulse train with single-pulse duration (FWHM) ~ $\tau_f=\SI{10}{fs}$ and period ~ $T_f=\SI{34}{fs}$, corresponding to a repetition frequency $\SI{30.3}{THz}$). The backward emission exhibits a longer pulse period, $T_b=47\;\mathrm{fs}$. This difference follows from the emission geometry: {in} the backward direction, the light pulses propagate against the electron's motion. Consequently, the total {envelope duration} is longer for backward radiation ($\Delta \tau_b=310\;\mathrm{fs}$) than for forward radiation ($\Delta \tau_f=243\;\mathrm{fs}$).

The temporal structure of emission can be tailored by tuning the incident electron energy. For instance, increasing the energy from \SI{1}{keV} to \SI{10}{keV} compresses both the single-pulse width (from \SI{10}{fs} to \SI{4.3}{fs}) and the pulse train period (from \SI{34}{fs} to \SI{7.5}{fs}), as shown in Fig. 4(b). The electron energy also strongly affects the angular distribution. The \SI{10}{keV} electrons produce a much more directional emission due to the relativistic effect. The forward-to-backward peak intensity ratio increases from $\sim$1.5 at \SI{1}{keV} to $\sim$9 at \SI{10}{keV} case. These results highlight the NU's potential as {an} on-chip source capable of generating highly structured femtosecond pulse trains with controllable repetition rate, wavelength, and directionality.

In conclusion, we propose a nanoscale, coherent, free-electron-driven light source based on a NU. The quantum mechanical analysis identifies key physical mechanism: the transverse near field drives a periodic oscillation of the electron wave packet while simultaneously providing a harmonic potential that coherently squeezes it. This squeezing counteracts natural dispersion and enables efficient long-range interaction. The NU platform offers opportunities for miniaturized lasers and integrated photonics \cite{zhangCoherentFreeelectronLight2023}, and the transverse modulation—particularly quantum squeezing—opens alternative routes for spatiotemporal wavefunction shaping \cite{tianFemtosecondlaserdrivenWireguidedHelical2017b} and for high-resolution imaging with spatially structured electron beams \cite{kruitDesignsQuantumElectron2016,priebeAttosecondElectronPulse2017}.

\begin{acknowledgments}
This work was supported by the National Key R$\&$D Program (No.2022YFA1604301 and No.2023YFA1406803) and Natural Science Foundation of China (No.12334013, No.124B2077, and No.92250306).
\end{acknowledgments}

\bibliographystyle{apsrev4-2}
\bibliography{ref}

\end{document}